\documentclass{emulateapj}

\usepackage[nointegrals]{wasysym}
\usepackage{amsmath}
\usepackage{amssymb}
\usepackage{alltt}
\usepackage{bm}
\usepackage{natbib}
\usepackage{ulem}
\usepackage{url}
\usepackage{graphicx}
\usepackage{listings}

\newcommand{\upstream}{upstream/unshocked }
\newcommand{\downstream}{downstream/shocked }
\newcommand{\rke}{RKE\ }
\begin{document}

\title{All Curled Up: A Numerical Investigation of Shock-Bubble
  Interactions and the Role of Vortices in Heating Galaxy Clusters}
\shorttitle{Shock-Bubble Interactions in Clusters}

\author{Samuel H.~Friedman and Sebastian Heinz}
\affil{Department of Astronomy, University of
  Wisconsin-Madison, 475 N.~Charter St., Madison, WI~53706}
\author{Eugene Churazov}
\affil{Max-Planck-Institute for Astrophysics,
  Karl-Schwarzschild-Str.~1, 85741 Garching, Germany}

\begin{abstract}

  Jets from active galactic nuclei in the centers of galaxy clusters
  inflate cavities of low density relativistic plasma and drive shock
  and sound waves into the intracluster medium.  When these waves
  overrun previously inflated cavities, they form a differentially
  rotating vortex through the Richtmyer-Meshkov instability.  The
  dissipation of energy captured in the vortex can contribute to the
  feedback of energy into the atmospheres of cool core clusters.
  Using a series of hydrodynamic simulations we investigate the
  efficiency of this process: we calculate the kinetic energy in the
  vortex by decomposing the velocity field into its irrotational and
  solenoidal parts.  Compared to the two-dimensional case, the
  3-dimensional Richtmyer-Meshkov instability is about a factor of 2
  more efficient.  The energy in the vortex field for weak shocks is
  $E_{vortex}\approx \rho_{ICM}\Delta\,v_{shock}^2V_{bubble}$ (with
  dependence on the geometry, density contrast, and shock width). For
  strong shocks, the vortex becomes dynamically unstable, quickly
  dissipating its energy via a turbulent cascade.  We derive a number
  of diagnostics for observations and laboratory experiments of
  shock-bubble interactions, like the shock-vortex standoff distance,
  which can be used to derive lower limits on the Mach number.  The
  differential rotation of the vortex field leads to viscous
  dissipation, which is sufficiently efficient to react to cluster
  cooling and to dissipate the vortex energy within the cooling radius
  of the cluster for a reasonable range of vortex parameters.  For
  sufficiently large filling factors (of order a few percent or
  larger), this process could thus contribute significantly to AGN
  feedback in galaxy clusters.
\end{abstract}

\section{Introduction}
While the gaseous atmospheres of galaxy clusters are virialized, they
are often far from relaxed.  Cooling gas can condense, form stars, and
funnel down to the central galaxy to feed the supermassive black hole
in its center.  Gas can enter the intra-cluster medium (ICM) through
accretion from cosmic filaments, ram pressure stripping of galaxies,
jets emanating from active galactic nuclei (AGN), and galactic
outflows due to supernova explosions.

The ICM has characteristic temperatures $\gtrsim 2$ keV and central
densities of order $\lesssim 0.01\,{\rm cm}^{-3}$, implying that the
cooling time of the central cluster gas in many clusters is much
shorter than the Hubble time.

However, observations with {\em XMM-Newton} and the {\em Chandra}
X-ray telescope (CXO) over the past decade have shown that cooling to
temperatures below about 1.5 keV is suppressed.  In order for the gas
in the ICM to maintain its temperatures over time, energy must be
added to counterbalance the radiative energy losses.

Extragalactic jets from AGN can provide this counterbalance due to the
very large energies injected into the ICM.
\citet{2007ARA&A..45..117M} provide an excellent overview of the
current paradigm of AGN feedback.

It is now well established that jet feedback happens through inflation
of X-ray cavities \citep[e.g.,][and references
therein]{churazov:00,mcnamara:00,fabian:00,Blanton:01,heinz:02,rafferty:06}.
At a minimum, the jets have to supply enough energy to create the
enthalpy associated with the radio bubbles / X-ray cavities they
inflate in the ICM by pushing aside the ICM and filling these cavities
with synchrotron emitting particles.

The estimates of the enthalpy ($4pV$) from observed cavities suggest
that AGN can indeed provide sufficient energy, though much of this
energy has a non-thermal form: The bubbles store up to 75\% of it as
non-thermal internal energy.  Depending on the inflation dynamics, an
appreciable fraction of the remaining energy can go into sound or
shock waves.  In order to understand how feedback works as a process,
it is important to understand how much of this energy is thermalized,
and where.

AGN are among the most highly variable phenomena, and jets in
particular are well known for being non-stationary.  It has been
suggested that observations of multiple generations of cavities and
successive concentric shock and sound waves in many clusters are
evidence for strong variability on duty-cycle time scales of order
1-10 million years, though dynamic instabilities, buoyancy, and shear
within the jets themselves can also account for some of these
observations \citep{morsony:10,soker:09}.

A detached cavity/bubble will buoyantly rise through the ICM, moving
further away from the central AGN, while the AGN creates a new bubble
and an associated outgoing shock wave.  While individual clusters only
afford us a snapshot view of this process, the multiple generations of
bubbles and waves observed in a number of nearby clusters, and the
high frequency of bubbles in cool core clusters suggest that this
process is constantly ongoing, filling the cluster with a spectrum of
bubbles and waves \citep{begelman:01}.

An increasing body of observations provides direct evidence of weak
shocks and non-linear sound waves in clusters with central radio
galaxies \citep{2006MNRAS.366..417F, forman:07, 2008MNRAS.386..278G,
  2009ApJ...697L..95B, 2010arXiv1006.5484M}.  As they propagate
outward, these waves must interact with previously inflated bubbles.
This interaction and the possible effect it can have on the
thermodynamics of the cluster is the subject of this paper.

\begin{figure}[tpb]
  \centering
  \includegraphics[width=0.95\columnwidth, clip=true, trim= 0in 0.5in 0in
  0.5in ]{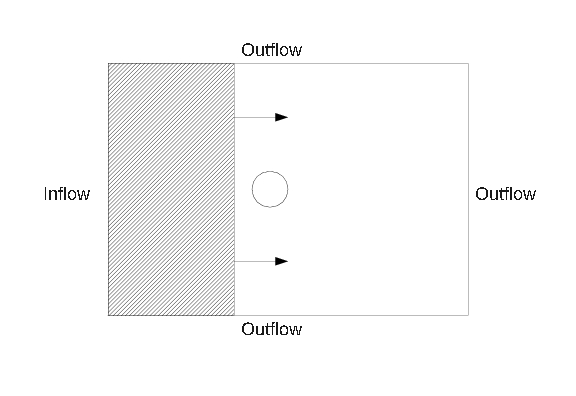}
  \caption{Boundary Conditions.  $x$-axis has mixed inflow/outflow
    boundary conditions, while the $y$- and $z$-axes have outflow
    boundary conditions.}
  \label{fig:setup_bcs}
\end{figure}

\citet{2005ApJ...634L.141H} (henceforth abbreviated as HC05) provided
a two dimensional study of shock--bubble interactions.  While sound
waves and weak shocks themselves have low efficiencies for
(non-adiabatic) dissipation (sound waves are completely
non-dissipative in the absence of viscosity or thermal conduction),
HC05 suggested that the dynamics of the
interaction of waves with previously inflated bubbles could extract
energy from the shock/sound wave and transform it to heat.

This potential heating mechanism for the ICM derives from the
Richtmyer-Meshkov Instability (RMI)
\citep{richtmyer:60,meshkov:69,brouillette_rmi:02} which operates when
a shock encounters a curved interface between two fluids.  The
proposed mechanism could work in concert with other forms of ICM
heating by shock waves and bubbles previously proposed in the
literature \citep[e.g.,][]{2005Natur.433...45M,2006ApJ...638..659M}.

The interaction of a shock wave with underdense bubbles has been
investigated in a broad body of work in the broader fluid dynamical
community, both experimentally and theoretically
\citep{1988JFM...189...23P, 1996JFM...318..129Q, 2001ShWav..11..209B,
  2005PhFl...17b8103L, 2006PhFl...18c6102G, 2007PhRvL..98b4502R,
  2008PhFl...20c6101R, 2008PhST..132a4020R, 2008JFM...594...85N,
  2009PhFl...21g4102L}.  However, none of these papers discuss the
prospect of the ensuing visco-rotational heating, and the extraction
of rotational kinetic energy suggested by
HC05.

Many previous studies examined the RMI through the use of 2D
simulations (e.g.\ \citet{1988JFM...189...23P}).  Such a treatment
necessarily excludes the impact of any movement of material through
the plane of simulation, eliminating some important aspects of
vorticity generation, as we will show in this paper.

\begin{figure}[tpb]
  \centering
  \includegraphics[width=0.95\columnwidth]{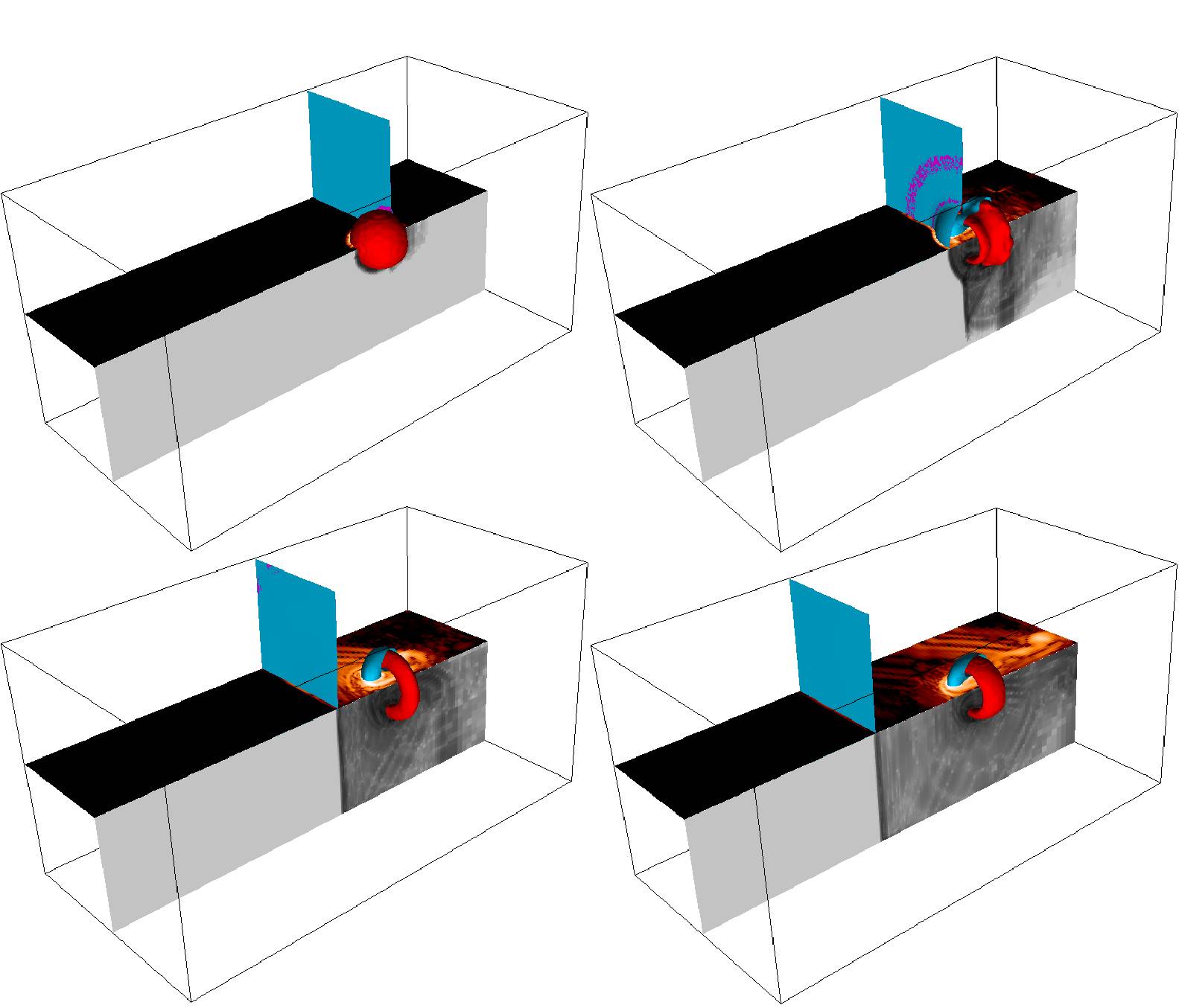}
  \caption{3D renderings of the shock--bubble interaction.  The short
    vertical plane that moves across the box is an iso-pressure
    contour that represents the shock front; the red surface is an
    iso-density contour that shows the under-dense sphere/vortex ring;
    the long (light-grey) vertical plane shows a cut through the
    baroclinic term, $|\nabla p \times \nabla \rho|$, and the long
    (dark-grey) horizontal plane show a cut through the enstrophy
    $\rho \omega^2$ (a proxy for the energy in the rotational velocity
    field).  Behind the blue-gree pressure contour, we show a magenta
    iso-density contour.  In locations where the two surfaces cross,
    the density and pressure gradient are mis-aligned, indication
    locations where vorticity is generated.}
  \label{fig:render}
\end{figure}

Returning to the astrophysical context, other studies have followed
the evolution of X-ray cavities / radio bubbles under the influence of
{\em buoyancy}, using both hydrodynamic and MHD simulation,
\citep{churazov:01,2004MPLA...19.2317G, 2005MNRAS.357..242R,
  2005ApJ...624..586J, 2007MNRAS.378..662R, 2008ApJ...684L..57L,
  2008MNRAS.389L..13S, 2008MNRAS.384.1377P, 2009ApJ...694.1317O,
  2009ApJ...704.1309D}.  

The effect of shock-bubble interaction in the context of radio relics
in the outer cluster was considered by \citet{ensslin:02}.  However,
outside of \citet{2005ApJ...634L.141H}, none have investigated the
effects of shock traversal of a cavity in the core of a cluster, and
the associated transfer of energy into the vortex.

\begin{table*}[tbp]
  \centering
  \begin{tabular}{|*{4}{l|}}
    \hline
    Name & Size [$R_{\rm bub}$] & Properties \\ \hline \hline
    Varying density & 16x8x8 & Density Ratios 
    $\rho_{\text{ICM}} / \rho_{\text{bub}}$ of 50, 20, 10, 5, \& 2 \\
    Varying Mach Numbers & 64x8x8 & Mach 1.02, 1.07, 1.25, 1.5, 1.75,
    2*, 4*, \& 8* \\ 
    Varying Mach (2.5D) & 128x8 & Mach 2, 4, \& 8 \\
    Multiple Bubbles & 64x16x16 & Impact Parameter of $0, 1, 2, 3, 4
    \; R_{\text{bub}}$ \\ 
    Multiple Bubbles & 64x8x8 & 3 Families of 2, 4, \& 8 bubbles \\
    Top-hat widths & 16x8x8 & Widths of $\lambda = 0.5, 0.6, 1,
    1.2, 2 \; R_{\text{bub}}$\\
    \hline
  \end{tabular}
  \caption{List of simulation parameters.  All simulations, except for
    the starred ones, have a maximum effective refinement of 5 levels
    (corresponding to a resolution of 32 cells across a bubble
    diameter).  The starred simulations, Mach 2, 4, and 8, have a maximum
    refinements of 7, 6, and 7, respectively.  Simulations 
    indicated as 2.5D are axi-symmetric.  Box sizes are given in units
    of bubble radii (the simulations are scale free).}
  \label{tab:table_of_simulations}
\end{table*}

The observational appearance of shock-generated vortices in clusters
was described in a companion paper \citep{heinz:11}, showing that,
while the interaction of waves and cavities in clusters is
unavoidable, a direct detection with X-ray methods will be difficult
(the so-called ``radio ear'' in the Virgo cluster might be the best
example of a vortex ring, given that it is trailing relatively closely
behind a moderate Mach number shock - see \citealt{owen:00,heinz:11}).

In this paper, we extend the study of HC05 to 3D, derive conditions
under which the RMI can effectively dissipate wave energy and heat a
cluster on a time scale shorter than the cooling time, and compare our
results to the general body of work on the RMI beyond the context of
clusters, both in the laboratory and in theory and simulation. We
ignore conduction, viscosity, gravity, and magnetic fields in the
simulations themselves, but we extend our discussion to an estimate of
the viscous dissipation rate of cluster-scale vortices.  We also limit
ourselves to the case of underdense bubbles.  For a study of the
inverse density scenario of shock-cloud interactions, we refer the
reader to, for example, \citet{1994ApJ...420..213K}.

In \S\ref{sec:methods}, we describe the methods used for our setup and
analysis.  Section \ref{sec:results} discusses our results,
\S\ref{sec:dissipation} discusses dissipation in the ICM, and
\S\ref{sec:conclusions} presents our conclusions.

\section{Methods}
\label{sec:methods}
We performed a large set of 3D simulations, spanning a range in
important simulation parameters.  Before describing the results in
general, we will first describe our simulation setup in
\S\ref{sec:mach2_bubble} and our analysis methods in
\S\ref{sec:rotvel_extraction}.  For reference, our fiducial simulation
is a Mach 2 shock running over an underdense bubble.  We will use this
simulation as a baseline to compare to other cases and to make the
connection to HC05.

We present a description and list of our entire set of simulations and
describe in \S\ref{sec:sim_table}, followed by a discussion of the
results.

\subsection{Simulation setup}
\label{sec:mach2_bubble}
We use the publicly available hydro code FLASH 3.3
\citep{2000ApJS..131..273F,2008PhST..132a4046D} which solves the
hydrodynamic equations on an adaptive Eulerian grid (in our case,
without gravity):
\begin{eqnarray}
  \frac{\partial \rho}{\partial t} +
  \nabla\cdot\left(\rho\mathbf{v}\right) & = & 0 \\
  \frac{\partial \rho\mathbf{v}}{\partial t} +
  \nabla\cdot\left(\rho\mathbf{v}\mathbf{v}\right) + \nabla P & = & 0
  \\
  \frac{\partial \left(\epsilon +
      \rho \left|\mathbf{v}\right|^2/2\right)}
  {\partial t} + \ \ \ \ & & \nonumber \\
  \nabla\cdot\left[\left(\epsilon + \rho\left|\mathbf{v}\right|^2/2 +
      P\right) \mathbf{v}\right] & = & 0
\end{eqnarray}
for the fluid density $\rho$, pressure $P$, velocity $\mathbf{v}$, and
internal energy density $\epsilon$.

FLASH employs the piece-wise parabolic method (PPM) to solve the
Riemann problem to second order
\citep{1984JCoPh..54..174C,1984JCoPh..54..115W}.  PPM uses a shock
capturing scheme to treat shocks.

We include a tracer fluid inside the bubble so that we can easily
track material initially inside the bubble as the simulation evolves.
Throughout the simulations presented in this paper, we use an
adiabatic equation of state, $P = \left(\gamma - 1\right) \epsilon$
with ratio of specific heats of $\gamma=5/3$.

The numerical setup used in this paper is similar to that described in
HC05: An underdense bubble (typically with density contrast 100 unless
otherwise indicated) is introduced to be at rest in, and in pressure
equilibrium with a uniform, stationary background medium.

We introduce a shock of Mach number $M$ traveling along the
x-direction downstream of the bubble such that it will overrun the
bubble in the course of the simulation.  The shock satisfies the
Rankine-Hugoniot jump conditions and sets the inflow boundary
conditions on the upstream x-axis to continually inject shocked gas
into the computational box, following the classic semi-infinite
shock-tube prescription.  We use outflow boundary conditions for the
remaining faces of the grid (see the sketch of the simulation setup in
Fig~\ref{fig:setup_bcs}).

We set the size of of the $y$- and $z$-axes equal to $8
R_{\text{bub}}$ and we vary the length of the $x$-axis to ensure the
relevant dynamics of the bubble are captured within the simulation
box, with a minimum value of $8 R_{\text{bub}}$.  Unless otherwise
noted, we used a maximum refinement level of 5, corresponding to a
resolution of 32 grid points across the bubble diameter and an
effective grid size of $(256+) \times 128 \times 128$.  We used
density as our refinement variable in order to refine both on the
shock front and the bubble.

For translation into physical units, a gas density of $n = 0.01\,{\rm
  cm^{-3}}$, a bubble radius of $R = 10\,{\rm kpc}$, and a gas
temperature of $T = 5 keV$ give typical box sizes of $80\,{\rm kpc}$
on a side and lengths from $240\,{\rm kpc}$ upward and characteristic
time (the sound crossing time of the bubble) of $1.8 \times
10^{7}\,{\rm yrs}$.

In this investigation, we vary the shock Mach number, the shape of the
bubble (we consider spherical and cylindrical bubble), the density
contrast, and the number of bubbles, and also investigate the effect
of changing the width of the shock (from semi-infinite to finite).

Because we were aiming to extend the 2D study from HC05 to three
dimensions, and because we are aiming to study a very basic
hydrodynamic scenario using a new analysis tool (Helmholtz
decomposition of the velocity field), we specifically kept the
numerical setup very simply, as described above: We did not include
gravity in the simulations, as we wished to investigate the efficiency
of the RMI, not of the buoyancy instability.  We also excluded
magnetic fields from this particular investigation, and we kept the
simulations as inviscid as possible, given the level of numerical
viscosity at the resolution allowed by our computational limitations,
i.e., we did not include a prescription for fluid viscosity in the
code.

Most of the simulations were run in-house on a 72 node Beowulf
cluster, typically using 64 cores and 128 GBytes of ram.  We estimate
a total runtime of 150,000 CPU hours, including analysis.

\begin{table*}
  \centering
  \begin{tabular}{|*{4}{l|}}
    \hline
    Name & Size [$R_{\rm cyl}$] & Properties \\ \hline \hline
    Varying Mach Numbers & Varied & Mach 1.02, 1.07, 1.25, 1.5, 1.75,
    2, 4, \& 8 \\
    Varying Angles & 32x16x16 & $L = 4 \; R_{\text{bub}}$, Angles = 0,
    30, 60, 90 degrees \\ 
    Varying Lengths & 256x(16/32)$^2$ & $x$-axis: $1, 2, 4, 8 \;
    R_{\text{bub}}$ \\ 
    Varying Lengths & 256x16x16 & $y$-axis: $1,2,4,8 \;
    R_{\text{bub}}$\\
    \hline
  \end{tabular}
  \caption{List of simulation parameters for shock-cylinder
    interactions.  All simulations have a maximum effective refinement
    of 5 levels.  sizes are measured in cylinder radii.  For the 
    various Mach numbers, some simulations have shorter sides parallel
    to the cylinder axis and use periodic boundary conditions.
    Simulations of cylinders with axes parallel to the $x$-axis (i.e.,
    the shock normal), the box size in the $y$
    and $z$ directions is set to 32 cylinder radii to ensure we
    contained all of the relevant dynamics within 
    our simulation volume.}
  \label{tab:table_of_simulations_cylinders}
\end{table*}

\subsection{Velocity decomposition}
\label{sec:rotvel_extraction}
Throughout this paper we will analyze simulations with the aim of
characterizing the energy extracted from the passing shock wave and
deposited in the vortex, following HC05.

We use Helmholtz's theorem to split up a 3D velocity field into two
components: an irrotational part and a rotational (solenoidal) part.
We can thus write the velocity field as:
\begin{align}
  \mathbf{v} &= \mathbf{v}_I + \mathbf{v}_R & \nabla \times
  \mathbf{v}_I &= 0 & \nabla \cdot \mathbf{v}_R &= 0
\end{align}

In essence, this method of extracting the rotational velocity and,
consequently, the rotational kinetic energy (RKE), is equivalent to
divergence cleaning of the 3D velocity field
\citep{1998ApJS..116..133B}.

While we use discrete Fourier transforms for this decomposition (which
is strictly only applicable in the case of periodic boundary
conditions), we show in a separate paper \citep{friedman_poisson} that
this technique provides an excellent approximation of the rotational
velocity field in cases where the vorticity, $\omega \equiv \nabla
\times \mathbf{v}$ vanishes near the edges of the simulation.  To
ensure that we satisfy this condition we use a sufficiently large grid
around the bubble/vortex.

\subsection{Rotational kinetic energy}
\label{sec:rke_intro}
Following HC05, it is straight forward to
estimate the scale on which the RMI can extract energy from the
passing wave.  For a bubble overrun by a shock with velocity jump
$\Delta v=v_{1}-v_{2}$, we can express the kinetic energy density of
the density perturbation, seen from the {\em \downstream} frame, as
$e_{\rm kin} = (\rho_{1}-\rho_{\rm bubble}) \Delta v^2/2 \sim
\rho_{1}\Delta v^2/2$.  For a bubble with an initial volume of $V_{\rm
  bubble}$, the fiducial energy scale of the problem is therefore
simply
\begin{equation}
  \label{eqn:fidual_energy}
  E_{\rm fid} = e_{\rm kin}V_{\rm bubble} = \rho_{1}\frac{\Delta
    v^2}{2}V_{\rm bubble}
\end{equation}

After measuring the actual extracted kinetic energy $E_{\rm rot}$ in
the vortex in the downstream frame as
\begin{equation}
  \label{eqn:e_rot_def}
  E_{\rm rot}=\int dV\,\rho\frac{v_{\rm rot}^2}{2}
\end{equation}
from our simulation, we can define an efficiency factor, $g_{2}$, as
\begin{equation}
  \label{eqn:define_g}
  g_{2} \equiv \frac{E_{\text{rot}}}{E_{\text{fid}}} =
  \frac{E_{\text{rot}}}{\frac{1}{2} V_{\text{bubble}} \rho_{1} \Delta v
    ^2 }
\end{equation}

Given the shift to the rest frame to the \upstream material, we have:
\begin{align}
  \Delta v & \equiv v_{1} - v_{2} = c_s M \left[ \frac{
      \frac{2}{M^2} + \left(\gamma - 1\right)}{\gamma + 1} -
    1 \right]
\label{eqn:v_2_definition}
\end{align}

We can derive an alternative representation for $g$ by expressing the
energy density of the approaching shocked gas in the {\em upstream}
material,
\begin{align}
  \label{eqn:define_g1}
  g_{1} &\equiv \frac{E_{\text{rot}}}{\frac{1}{2} V_{\text{bubble}} \rho_2
    {\Delta v}^2 } \\ \nonumber
  &= g_{2}\frac{\left(\gamma + 1\right) +
    \left(\gamma - 1\right)\left(M^2 -
      1\right)}{\left(\gamma + 1\right)M^2} < g_{2}
\end{align}

$g_1$ is associated with $\rho_2$, the \downstream density, and $g_2$
is associated with $\rho_1$, the \upstream density.  In other words,
$g_{2}$ measures the bubble energy from the point of view of the
shocked, \downstream medium, compared to the kinetic energy missing
from the evacuated bubble volume approaching the shock.  $g_1$
represents the ratio of the rotational kinetic energy to the kinetic
energy in the \downstream volume contained within the bubble volume
(i.e., from the point of view of the \upstream, unshocked medium).  We
will use $g_{1}$ throughout the rest of the paper unless otherwise
noted, with easy conversion given eq.~(\ref{eqn:define_g1}).

\subsubsection{Vorticity Evolution}
To illuminate the development of the RMI, it is instructive to
consider the vorticity equation (the curl of the Euler equation):
\begin{align}
\label{eqn:curl_euler_eqn_early}
\frac{D \mathbf{\omega}}{D t} &= \underbrace{(\mathbf{\omega} \cdot
  \mathbf{\nabla}) \mathbf{v}}_{\text{Vortex Stretching}} -
\underbrace{\mathbf{\omega} (\mathbf{\nabla} \cdot
  \mathbf{v})}_{\text{Vortex Compression}} \\\nonumber &\quad +
\underbrace{\frac{1}{\rho^2} (\nabla \rho \times \nabla
  p)}_{\text{Baroclinic}}
\end{align}
The last term in \ref{eqn:curl_euler_eqn_early} is commonly referred
to as the baroclinic term, and it is present in both 2D and 3D.
While the vortex compression term is present in both 2D and 3D
simulations, the vortex stretching term is present {\em only} in the
3D case.  It represents $m\geq 1$ modes, since for the $m=0$ mode the
$\phi$-derivatives vanish (this is the axi-symmetric 2.5D case described
in \S\ref{sec:sim_table}).

\subsection{Simulation table}
\label{sec:sim_table}
Using our fiducial run as a baseline, we modified the simulations by
changing the number of bubbles, the Mach number, the width of shocked
material ($\lambda$), the bubble density, and the bubble geometry
relative to the shock front.  The grid of simulations discussed in the
following is laid out in tables \ref{tab:table_of_simulations} and
\ref{tab:table_of_simulations_cylinders}.

\begin{figure}[tpb]
  \centering
  \includegraphics[width=0.48\columnwidth]{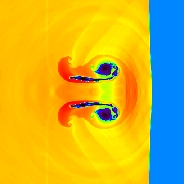}
  \includegraphics[width=0.48\columnwidth]{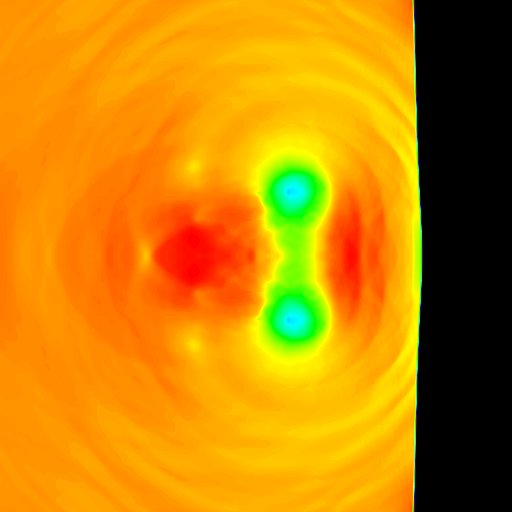}
  \includegraphics[width=0.48\columnwidth]{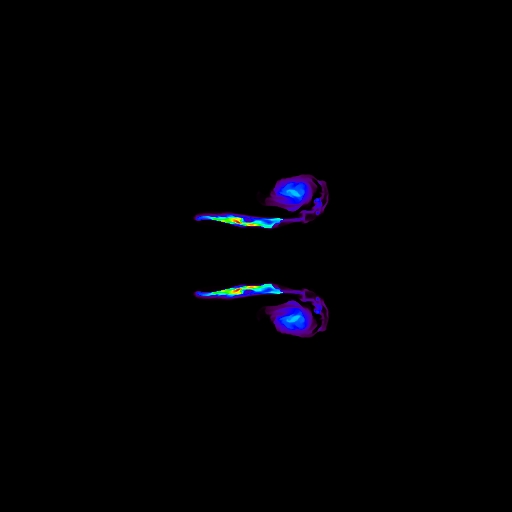}
  \includegraphics[width=0.48\columnwidth]{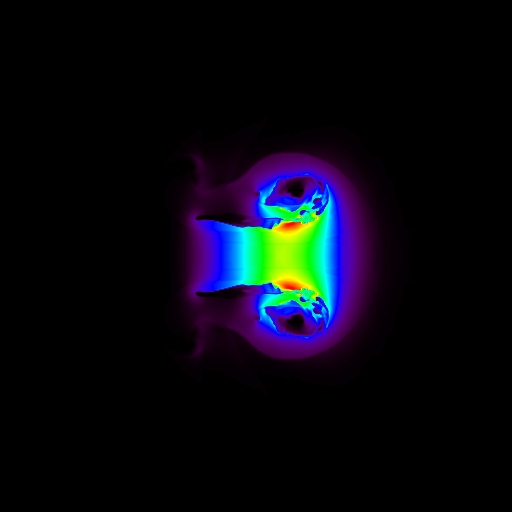}
  \caption{2D slices through the center of the simulations box show
    density (top left), pressure (top right), tracer fluid mass
    fraction (bottom left), and rotational kinetic energy (bottom
    right) for a Mach 2 shock, spherical bubble, simulated in a
    uniform grid ($1024 \times 512 \times 512$) grid.  The acoustic
    velocity field is visible as roughly spherical outgoing waves,
    while the primary and secondary vortex rings (PVR and SVR,
    respectively) are visible as dark, low-density enclosures in the
    density.}
  \label{fig:dens_slice_0030_nz-2.jpg} 
\end{figure}

In simulations with cylindrical bubbles instead of spheres, we label
cylinders with axes {\em perpendicular} to the shock normal
``infinite'' cylinders because they represent a 3D extension of the
plane-parallel 2D circular bubble in
HC05.  The main families of simulations
investigate the dependence of $g$ on fundamental parameters of the
hydrodynamics: Mach number, density ratio of bubble to environment,
and shock thickness.

Due to the strong long--term time dependence we find in the case of
high Mach number shocks (see Fig.~\ref{fig:g_vs_t_mach4-8_max20}) that
was absent in the 2D simulations of HC05,
we decided to investigate a 2.5 dimensional (axi-symmetric) analog of
the three dimensional spherical bubble case.  Using cylindrical
coordinates, $(r,z)$, and placing the center of the bubble at $r=0$,
we reproduced the set of simulations of spherical bubbles for
different Mach numbers.  To calculate the \rke we then mapped the 2.5D
cylindrical velocities onto a 3D Cartesian grid and treated them the
same way as we treat our 3D regular simulations.

Throughout most of our simulations we use a shock driven by a
semi-infinite piston, i.e., a shock with thickness $\lambda \gg
R_{\text{bub}}$.  As shown in HC05, the
efficiency of the RMI depends on $\lambda$ when $\lambda \sim R$.
Extending this test to 3D, we simulate shocks with finite width.  The
resulting shock has a top-hat structure at $t=0$.  At later times, the
back end of the shock expands into a rarefaction wave.

\subsection{Description of bulk dynamical properties}

The general evolution of a bubble subject to the RMI has been
described in numerous publications.  In the context of clusters, we
refer the reader to HC05 and \citep{ensslin:02}.  We will only briefly
review the evolution, illustrating a few noteworthy points using
rendered images from our simulations.

\subsubsection{Fiducial Run}
Fig.~\ref{fig:render} shows the interaction of a single spherical
bubble with a Mach 2 shock, our fiducial case.  As expected, the shock
enters the bubble and the pressure increase quickly and traverses the
bubble, in effect ``smearing out'' the shock across the entire bubble
surface.  At the same time, the bubble has negligible inertial
density, implying that the shocked material travels quickly through
the bubble, which leads to the formation of a vortex ring such that
$\hat{\vec{r}}\cdot\left(\vec{\omega}\times\hat{\vec{n}}\right) > 0$,
where $\hat{\vec{r}}$ is the vector from the vortex axis to the vortex
ring, $\vec{\omega}$ is the vorticity inside the vortex, and
$\hat{\vec{n}}$ is the shock normal, aligned with the shock velocity.
Figure \ref{fig:render} shows the primary vortex ring as a red density
contour surface.

The figure also shows the magnitude of the baroclinic term and the
local enstrophy, $\rho\omega^2$ (related to the rotational kinetic
energy).  The interaction generates vorticity whenever the baroclinic
term in equation \ref{eqn:curl_euler_eqn_early} does not vanish.

Figure~\ref{fig:dens_slice_0030_nz-2.jpg} shows 2D slices through some
of the relevant fluid variables in our fiducial Mach 2 run.

\subsubsection{Non-Linear Vortex Interaction: Simulations of Multiple
  Bubbles}

While the simulation of a single spherical bubble presents the
cleanest possible numerical laboratory to study the efficiency of the
RMI, one of the questions we aimed to investigate is whether and how
$g_1$ depends on complications like non-sphericity of the bubble and
non-linear interaction of vortex rings from multiple neighboring
bubbles.

We designed two sets of numerical experiments to investigate the
interaction of multiple vortex rings and its effect on $g_1$.  To
facilitate comparison with other simulation series and with the
fiducial run, we simulated a Mach 2 shock.

\begin{figure}[tpb]
  \centering
  \includegraphics[width=0.95\columnwidth]{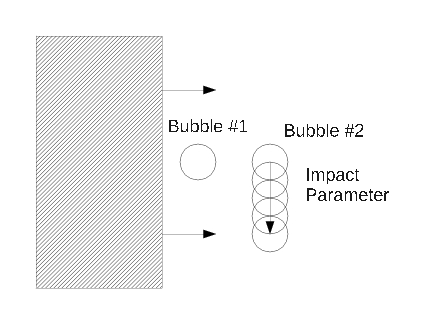}
  \caption{Multiple bubbles with two bubbles and a varying impact
    parameter.  We adjust the position of bubble \#2 in increments of
    $R_{\text{bub}}$ while keeping the position of bubble \#1 the
    same.}
  \label{fig:impact_parameter}
\end{figure}

The first approach places two bubbles offset from each other by $4 \;
R_{\text{bub}}$ in the $x$ direction (the shock-propagation direction)
and displaced in the $y$ direction (perpendicular to $v$) by integer
increments of $R_{\text{bub}}$ (see
Figure~\ref{fig:impact_parameter}).

\begin{figure}[tpb]
  \centering
  \includegraphics[width=0.95\columnwidth]{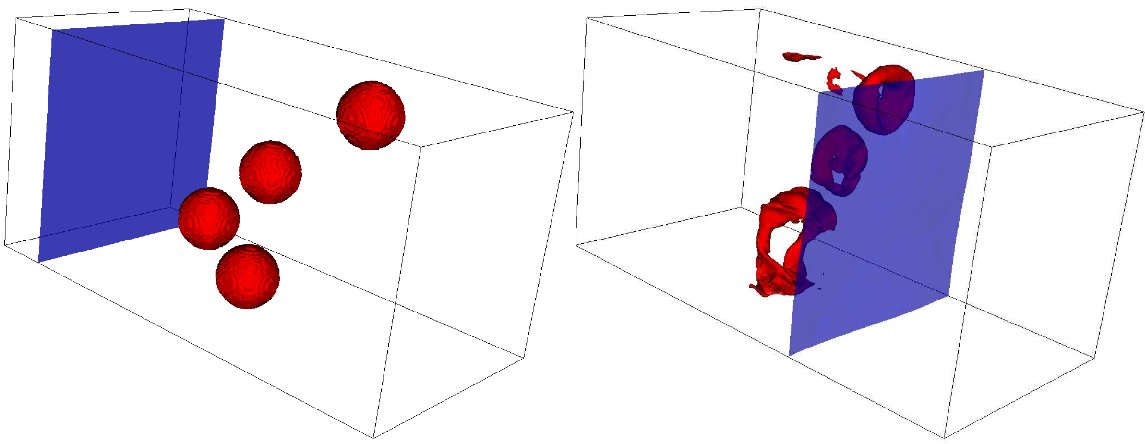}
  \caption{Multiple bubbles with four bubbles randomly placed.  At the
    far end of the box is a Mach 2 shock approaching the bubbles.}
  \label{fig:multi_bub_100810.jpg}
\end{figure}

The second approach simulates three families each of 2, 4, or 8
non-overlapping bubbles, randomly placed in a box of volume $8^3
R_{\text{bub}}^3$ with periodic boundary conditions along the $y$- and
$z$-axes.  Figure \ref{fig:multi_bub_100810.jpg} provides before and
after surface renderings of a simulation of a shock interacting with
four bubbles.

\subsubsection{Resolution Study}
\label{sec:resolutionstudy}

\begin{figure}[tpb]
  \centering
  \includegraphics[width=0.95\columnwidth]{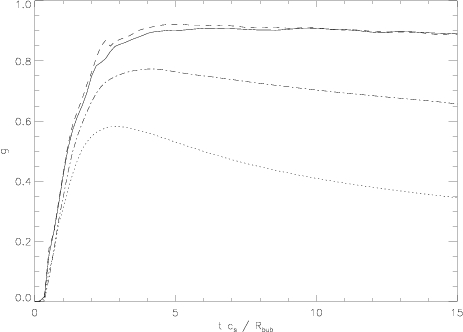}
  \caption{Resolution study for a Mach 2 sphere.  We represent maximum
    refinement levels of 3, 4, 5, and 6 with dotted, dot dashed, solid
    and dashed lines respectively.  Level 5 appears to have the
    minimum necessary resolution to refine the relevant dynamics,
    corresponding to a resolution of 32 cells across the bubble.}
  \label{fig:sph_resolution_study}
\end{figure}

To ensure that we have simulated the relevant dynamics and that
numerical resolution effects did not influence our results, we
performed a resolution study by increasing the refinement in our
adaptive mesh, the results of which are presented in Figure
\ref{fig:sph_resolution_study}.

It is clear from the figure that the value of $g$ determined from the
simulation converges at a maximum refinement level of 5 at the surface
of the bubble, i.e., at an effective resolution of 32 cells across the
bubble.  In other words, $g_{1}$ measured in our simulations is
robust.

A high-resolution 2.5D axi- symmetric simulation in a uniform grid also
shows excellent convergence with the 3D simulation at refinement level
5 and above, corresponding to a resolution of 32 cells across the
bubble, as can be seen in Figure \ref{fig:select_plot-11_100803.ps}.
The highest resolution simulations we ran of the Mach 2 case have an
effective resolution of 128 cells across the bubble, corresponding to
a numerical Reynolds number of order $Re \sim 10^4$.  For low Mach
numbers, we are therefore confident that our simulations fully resolve
the gross vortex dynamics and that the numerical values we determine
for $g_{1}$ are correct.

The situation changes at high Mach number: The highly turbulent and
fragment flow around the Mach 8 vortex shown in Figure
\ref{fig:bw_3d_mach8_bis_hdf5_plt_cnt_in0060.jpg} exhibits structure
all the way to the resolution limit and the vortex breaks apart (see
discussion in \S\ref{sec:high_mach_numbers} and
\ref{sec:conclusions}).

In terms of the evolution of $g_{1}$, our resolution study shows that
simulations at Mach numbers well above 2 have not converged, as might
be expected from the degree of turbulence present in the flow and we
limit discussion of these cases to a phenomenological description of
the dynamics observed in our simulations for the interested reader,
since higher resolution simulations would be computationally
unfeasible given reasonable resources.

%
%

\section{The Efficiency of the Richtmyer-Meshkov Instability in 3D}
\label{sec:results}

Before discussing the relevance of vortex creation to the thermal
evolution of galaxy clusters and AGN feedback, we will discuss our
simulation results in the context of traditional fluid dynamics and
compare them to experiments and previous theoretical and numerical
work.

\begin{figure}[tpb]
  \centering
  \includegraphics[width=0.95\columnwidth]{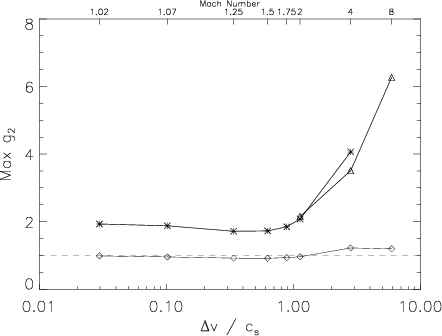}
  \caption{RMI efficiency factor $g_{2}$ as a function of Mach number.
    The stars represent spheres, diamonds represent ``infinite''
    cylinders, and triangles represent spheres in the 2.5D
    axi-symmetric simulations.  The dashed line represents the
    approximation given by HC05,
    corresponding to the kinetic energy contained in a bubble of
    volume $V_{\rm bub}$ in the \upstream medium, as seen from the
    \downstream frame.\label{fig:g_vs_v2_bub_cyl_unshocked}}
\end{figure}

\begin{figure}[tpb]
  \centering
  \includegraphics[width=0.95\columnwidth]{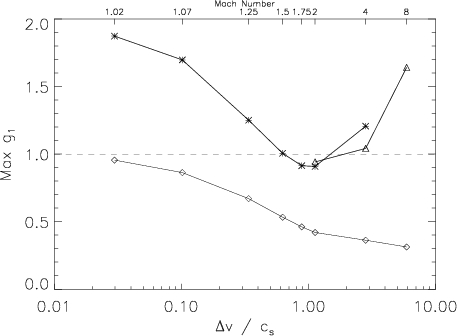}
  \caption{RMI efficiency factor $g_{1}$ as a function of Mach number.
    Same symbols and Mach numbers as those in Figure
    \ref{fig:g_vs_v2_bub_cyl}.   The dashed line represents the kinetic
    energy of a bubble with volume $V_{\text{bub}}$ in the \downstream
    material, as seen from the \upstream frame.
    \label{fig:g_vs_v2_bub_cyl}}
\end{figure}

Having introduced the \rke as a measure of the efficiency of the RMI
in \S\ref{sec:rke_intro}, we will first draw a comparison to the
previous 2D results from HC05 and then
discuss an extension of the investigation to a broader set of
questions in a general fluid mechanics context, such as the non-linear
interaction between vortices, and a comparison to previous studies.

\subsection{Low Mach numbers}
In their 2D investigation of the RMI,
HC05 showed that, in 2D, vortex creation
by the RMI can be surprisingly efficient, with $g_{2}\approx 1$ for
bubbles much smaller than the depth $\lambda$ of the shock, but large
compared to the shock thickness $\delta$.  The efficiency depends on
geometric factors, but HC05 showed that,
over a range of Mach numbers from $M=1.01$ to $M=4$, the efficiency
$g_{2}$ is independent of $M$.

However, as is well known in the case of other fluid processes, the
behavior can be qualitatively different in three dimensions.  The most
obvious difference between the 2D and the 3D case is the fact that a
spherical bubble has more surface area per volume, and a larger
fraction of that surface area is oriented perpendicular to the shock
normal.

Since vorticity is generated if and only if the baroclinic term
$\nabla p \times \nabla \rho \not= 0$ is non-vanishing, a larger
fraction of bubble surface misaligned with the shock means a larger
area over which the baroclinic term is non-vanishing.  One should
therefore expect a spherical (3D) bubble to have higher efficiency at
generating vorticity than an infinite (2D) cylinder.

This can easily be verified from
Fig.~\ref{fig:g_vs_v2_bub_cyl_unshocked} which plots $g_2$ as a
function of shock Mach number for spherical (3D) and cylindrical (2D)
bubbles.  Our 2D results reproduce the finding that $g_{2}=1$ for
cylinders from HC05.  For the spherical
case, we find that $g_{2}\approx 2$ for Mach numbers smaller than
$M<2$, confirming the significantly increased efficiency of the RMI in
3D.

For comparison, we have also plotted the peak value of $g_{1}$ in
Figure~\ref{fig:g_vs_v2_bub_cyl}.  As expected, the two curves
converge for low Mach numbers, where the shock compression ratio
approaches unity.

\subsection{High Mach numbers}
\label{sec:high_mach_numbers}
For larger $M$, a second clear difference from 2D emerges:
$g_{2}$ starts deviating significantly from a constant.  The {\em
  peak} value of $g_{2}$ is strongly increased over $g_{2}=2$ at high
Mach numbers.  This occurs both in the full 3D simulations as well as
in the 2.5D axi-symmetric simulations in the same geometry as the 3D
case (i.e., a spherical bubble, triangles in
Figures~\ref{fig:g_vs_v2_bub_cyl_unshocked}~and~\ref{fig:g_vs_v2_bub_cyl}).

As expected for higher Mach number shocks, the increase in post-shock
pressure implies that the vortex ring is becomes more strongly
compressed, leading to the formation of a very thin vortex ring.
Properly simulating the dynamics of such a ring requires a
significantly higher maximum refinement level than is the case at low
Mach numbers (see Table \ref{tab:table_of_simulations}) and ultimately
prohibited us from reaching convergence of our high Mach number
simulations.

\begin{figure}[tpb]
  \centering
  \includegraphics[width=0.95\columnwidth]{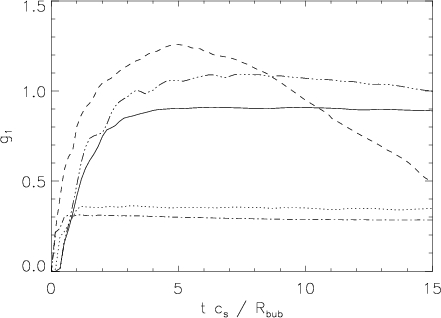}
  \caption{$g_{1}$ vs.~time for Mach 4 and Mach 8 simulations.  The
    solid line represents a Mach 2 spherical bubble, the dash dot
    dotted line represents a Mach 4 spherical bubble, the dotted line
    represents a Mach 4 ``infinite'' cylinder, the short dashed line
    represents a Mach 8 spherical bubble with a maximum refinement
    level of 6, and the long dashed line represents a Mach 8 spherical
    bubble with a maximum refinement level of 7, and the dash dotted
    line represents a Mach 8 ``infinite'' cylinder.}
\label{fig:g_vs_t_mach4-8_max20}
\end{figure}

\begin{figure}[tpb]
  \centering
  \includegraphics[width=0.95\columnwidth]{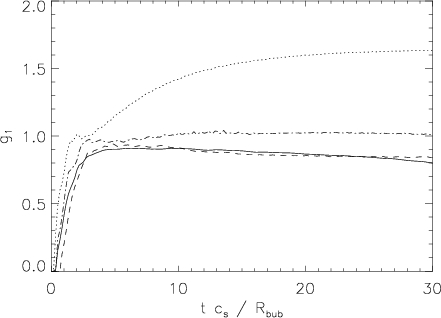}
  \caption{$g_1$ vs.~$t$ for spherical bubbles in a 2D axi-symmetric
    simulations.  The dashed line represents a Mach 2 shock, the dash
    dotted line a Mach 4 shock, and the dotted line a Mach 8 shock.
    The solid line represents a spherical bubble with a Mach 2 shock
    in 3D.}
\label{fig:select_plot-11_100803.ps}
\end{figure}

In addition, the long term evolution of the vortex also changes at
high Mach number, as shown in Figure~\ref{fig:g_vs_t_mach4-8_max20}.
While in the $M<2$ case $g_{1}$ remains essentially constant
for the entire duration of the simulation once the shock has crossed,
$g_{1}$ shows a measurable decline in the Mach 4 case and an even more
marked decline in the Mach 8 case, after reaching its significantly
{\em higher} peak value.  This behavior, however, is absent in the
2.5D case, as can be seen in the asymptotic behavior of $g_{1}$ in
Figure~\ref{fig:select_plot-11_100803.ps}.

\begin{figure}[tpb]
  \centering
  \includegraphics[width=0.95\columnwidth]{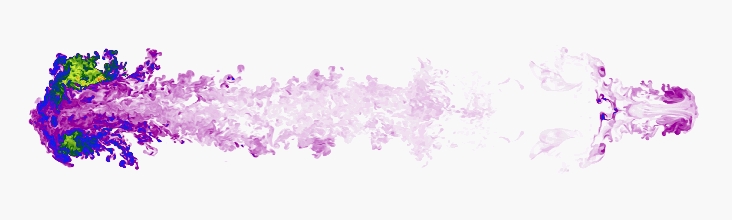}
  \caption{Tracer fluid distribution for a full 3D Mach 8 simulation
    at $t c_s / R_{\text{bub}} \approx 12$.  Note the high degree of
    complexity present in the flow, compared to the lower Mach number
    cases shown above.
    \label{fig:bw_3d_mach8_bis_hdf5_plt_cnt_in0060.jpg}}
\end{figure}

The reason for the rapid decline in ordered vortex energy is the
development of turbulence around the vortex, which completely disrupts
the vortex ring fairly shortly after the shock crossing.  This can be
seen in Fig.~\ref{fig:bw_3d_mach8_bis_hdf5_plt_cnt_in0060.jpg}, which
shows the distribution of tracer fluid initially contained inside the
bubble after $\sim 50$ shock crossing times.  The vortex is completely
disrupted and turbulence has cascaded down to the smallest resolved
scales.

As stated in \S\ref{sec:resolutionstudy}, our simulations are not
fully resolving the flow at the highest Mach numbers and we will
therefore defer a quantitative investigation of the efficiency of the
high-Mach number RMI to future work.

\subsection{The effects of geometry and non-linear vortex interactions
  on the RMI}

Filamentary relativistic plasma in cluster atmospheres (as well as
many other astrophysical objects) will often deviate from a purely
spherical geometry.  The increase in $g$ from 2D cylinders to 3D
spheres demonstrates the effect of geometry on the efficiency of the
RMI.  Following HC05, we evaluated the
dependence of $g$ on bubble aspect ratio and inclination.

\begin{figure}[tpb]
  \centering
  \includegraphics[width=0.95\columnwidth]{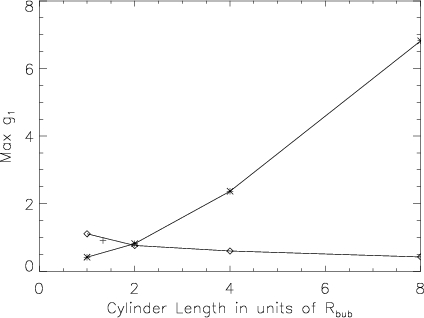}
  \caption{Efficiency $g_{1}$ of the RMI for cylindrical bubbles of
    varying length and orientation for a Mach 2 shock, plotted against
    cylinder aspect ratio (length in units of radius).  Cylinders
    perpendicular and parallel to the shock normal are shown as
    diamonds and stars, respectively. The cross indicates the value of
    $g_{1}$ for a spherical bubble.}
\label{fig:g_vs_cyl_length}
\end{figure}

Figure~\ref{fig:g_vs_cyl_length} shows that cylinders oriented
perpendicular to the shock normal are less efficient than spheres (as
already discussed above), and that cylinders oriented along the shock
normal are more efficient.  This makes intuitive sense, as vorticity
generation is maximized when the portion of the bubble surface
perpendicular to the pressure gradient is maximized.

\begin{figure}[tpb]
  \centering
  \includegraphics[width=0.95\columnwidth]{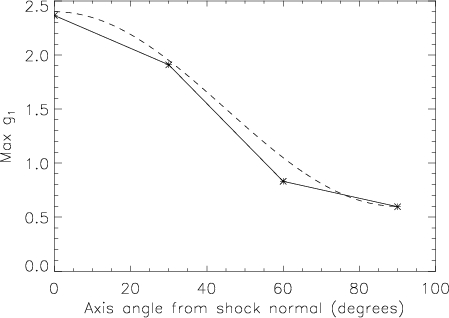}
  \caption{Maximum $g_1$ value vs.~axis angle of a cylinder from shock
    normal (degrees) for a Mach 2 shock.  Cylinders have an axial
    length $L=4R_{\text{bub}}$.  The dashed line is the curve given by
    $1.5+0.9 \times \cos(\theta \pi /90)$.}
\label{fig:g_vs_cyl_angle}
\end{figure}

It is reasonable to expect that $g$ should vary smoothly as we vary
the inclination with the shock normal.  This effect is shown in
Fig.~\ref{fig:g_vs_cyl_angle}, along with an simple, ad-hoc
parameterization as a sinusoid (in rough qualitative agreement with
the data).

If we suppose that a sinusoid provides a good approximation and that a
cylinder of aspect ratio $\mathcal{R}\gg 1$ has a minimum efficiency
of $g_{2,{\rm min}} \approx 1$ and a maximum efficiency of $g_{2,{\rm
    max}}(\mathcal{R})$, the {\em average} efficiency for a randomly
oriented set of filaments should be $\langle g_{2} \rangle \approx
0.5(0.5 + g_{2,{\rm max}})$.

The presence of multiple filaments or bubbles also implies that the
individual vortices created in the interaction with a passing wave
will interact with each other.  Individual vortex velocity profiles
can be expected to fall off roughly as $|v(r)| \propto r^{-3}$ with
distance $r$ from the center at large $r \gg R$, where $R$ is the
characteristic vortex size, \citep{batchelor,heinz:11}.  Two vortices
separated by a distance of order $R$ will interact strongly, while one
can expect the interaction to be insignificant for large distances,
given the steep dependence of $v$ or $r$.

\begin{figure}[tpb]
  \centering
  \includegraphics[width=0.95\columnwidth]{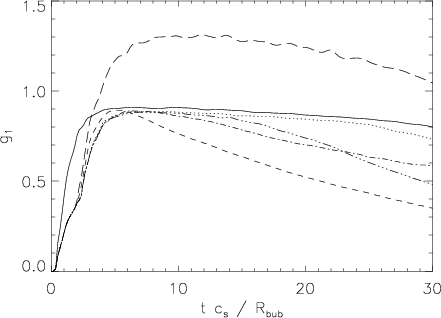}
  \caption{Efficiency $g_1$ for two interacting bubbles, offset along
    the shock normal by 4 bubble radii and transverse to the shock
    normal by an increasing impact parameter for a Mach 2 shock; {\em
      solid line}: single bubble; {\em long dashed}: two bubbles,
    transverse offset $b=0{\, R_{\text{bub}}}$; {\em short dashed}:
    $b=1\, R_{\text{bub}}$; {\em dash--dotted}: $b=2 \,
    R_{\text{bub}}$; {\em dashed--triple--dotted}: $b=3 \,
    R_{\text{bub}}$; {\em dotted}: $b=4 \,
    R_{\text{bub}}$.} \label{fig:g_vs_t_two_bubble_impact_linestyle}
\end{figure}

We attempted to quantify this interaction with two numerical
experiments.  Figure~\ref{fig:g_vs_t_two_bubble_impact_linestyle}
shows the results from the experiment described in
Figure~\ref{fig:impact_parameter}: two spherical bubbles placed behind
each other with a varying lateral offset.  Given the dependence of $g$
on cylinder inclination, one would expect an aligned configuration
(with zero lateral offset) to have higher efficiency than a
configuration in which two vortices are laterally offset by some
impact parameter $b > 0$.

This is supported by the simulations.  For two aligned bubbles (with
longitudinal offset of 4 bubble radii), the net efficiency is larger
than for two individual bubbles, while the net efficiency is reduced
for bubbles with a lateral offset.
Figure~\ref{fig:g_vs_t_two_bubble_impact_linestyle} shows the temporal
evolution of the interaction: As we reduce the impact parameter, the
efficiency decreases from that of two isolated vortices as the
inter-vortex interaction increases.

With increasing offset, the effect appears later and becomes weaker,
as expected.  At an impact parameter of 4 bubble radii, the result
becomes virtually identical to the isolated case.  This has
implications for the effect of bubble filling factor on the efficiency
$g$ of the RMI: For volume filling fractions below $\approx 2\%$, we
expect relatively little non-linear interaction between vortices
(corresponding roughly to the offset of 4 bubble radii both laterally
and longitudinally for which we measured little effect on $g$), while
for filling factors larger than about 2\%, we should expect a
measurable effect.

\begin{figure}[tpb]
  \centering
  \includegraphics[width=0.95\columnwidth]{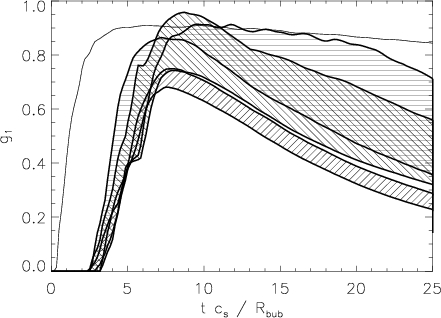}
  \caption{$g$ vs.~time for a Mach 2 shock with multiple, randomly
    placed bubbles (for the filling factors given below).  {\em
      Horizontally hatched}: the two bubbles; {\em upper-left to
      lower-right hatched:} four bubbles; {\em lower-left to
      upper-right:} eight bubbles, compared to the single bubble case
    (thin solid line).  The average filling factors for the different
    2, 4, and 8 bubble realizations are approximately 1.5\%, 3\%, and
    6\% respectively.}
  \label{fig:multi_bubble_hatched_families}
\end{figure}

Since, on average, two randomly placed vortices will be aligned at an
angle relative to the shock normal, this effect will reduce the
average efficiency $g$.  This can be seen in
Fig.~\ref{fig:multi_bubble_hatched_families}, which shows the results
of our simulations of ensembles of randomly placed bubbles.  The
hatched areas plotted show the envelope of $g$ spanned by the
different realizations for filling factors of 1.5\%, 3\%, and 6\% as a
function of time, compared to a single bubble.  The presence of
multiple bubbles reduces the peak efficiency and introduces a temporal
decline not present in the simulations of individual bubbles at the
same Mach number\footnote{In the case of only two bubbles, the range
  spanned by $g$ is largest, given that two bubbles close together can
  interact strongly and one would expect larger relative variance for
  a smaller number of bubbles.}.

\subsection{Secondary effects}

As discussed in HC05, the structure of the wave passing over a bubble
and the density contrast affect the efficiency of the RMI.  The bulk
of the simulations in this paper were carried out under the assumption
of having a small bubble radius compared to the pulse width of the
shock (i.e., we simulated the shock as a semi-infinite piston) and a
density contrast of 100.

\subsubsection{Shock geometry}

\begin{figure}[tpb]
  \centering
  \includegraphics[width=0.95\columnwidth]{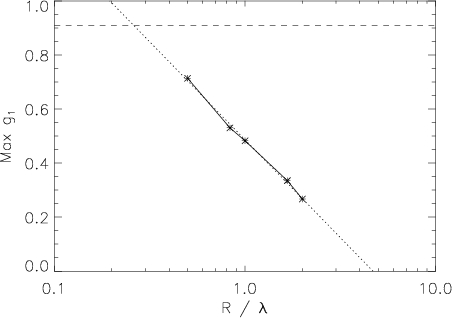}
  \caption{$g_1$ as a function of pulse width for a Mach 2 top-hat
    shock.  The dashed line represents a top-hat shock with infinite
    length or a infinitely long piston. The dotted line represents the
    fit to the data given by $g_1 = 0.49 - 0.31 \times \ln (
    {R_{\text{bub}}}/{\lambda} )$.}
  \label{fig:g_vs_shock_width}
\end{figure}

\begin{figure}[tpb]
  \centering
  \includegraphics[width=0.95\columnwidth]{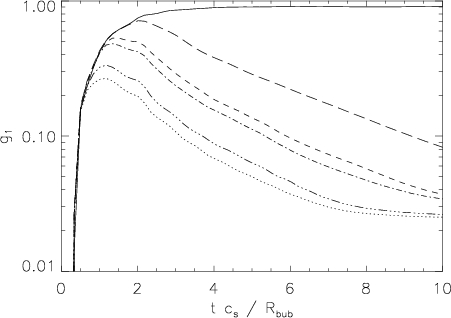}
  \caption{$g_1$ as a function of time for a Mach 2 top-hat shock of
    width $\lambda=2 R_{\text{bub}}$ (long dashed), $\lambda=1.2
    R_{\text{bub}}$ (short dashed), $\lambda=1.0 R_{\text{bub}}$
    (dash-dotted), $\lambda=0.6 R_{\text{bub}}$
    (dashed-triple-dotted), and $\lambda=0.5 R_{\text{bub}}$ (dotted).
    The solid line represents the semi-infinite
    piston. \label{fig:g_vs_t_vary_widths_linestyle}}
\end{figure}

To investigate the effect of finite shock width on $g$, we injected a
top-hat pressure and density perturbation (satisfying the shock jump
conditions at the leading edge) to travel through the grid, with
width\footnote{The {\rm thickness} of the shock will be of the order
  of the mean free path of the particles $\lambda_{\rm mfp} \sim pc$,
  and thus small compared to typically observed bubble sizes,
  justifying our approximation of the shock as a sharp
  discontinuity.}, $\lambda \leq 2R$ (for $\lambda \gg R$, we expect
the result of the semi-infinite piston to hold).  The results are
presented in
Figs.~\ref{fig:g_vs_shock_width}~and~\ref{fig:g_vs_t_vary_widths_linestyle}.

As in the case of the non-linear interaction between vortices, $g_1$
reaches a peak value and declines as the inverted pressure gradient at
the back of the perturbation reverses some of the vorticity generation
of the shock.  In the limit of narrow shocks (where the pulse width is
small compared to the bubble size) we should expect that the peak in
$g$ is significantly reduced compared to the semi-infinite piston
case, as the shock has signficantly reduced energy compared to the
maximum possible.  This is borne out by the results shown in
Fig.~\ref{fig:g_vs_shock_width}.  Similar to the 2D case presented in
HC05, smaller bubbles relative to the pulse width are more efficient
at generating vorticity.

\subsubsection{Density Contrast}

The X-ray cavities in the centers of cool core clusters discussed in
this paper are generally filled with radio synchrotron emitting
plasma.  While it is reasonable to assume that the density contrast
between the radio plasma and the thermal ICM gas is very large
(validating our choice of $\rho_{\rm bub} \ll \rho_{\rm ICM}$), the
best observational upper limits of the filling fraction of thermal gas
inside the bubbles are about an order or magnitude larger:
\cite{sanders:07} report an upper limit of 15\% on the filling factor
of thermal gas inside the inner cavities in the Perseus cluster at a
temperature of 10 keV or below, which translates into a limit of
$\rho_{\rm bub}/\rho_{\rm ICM}\leq 0.15$ assuming pressure equilibrium
between the cavity and the surrounding ICM.

\begin{figure}[tpb]
  \centering
  \includegraphics[width=0.95\columnwidth]{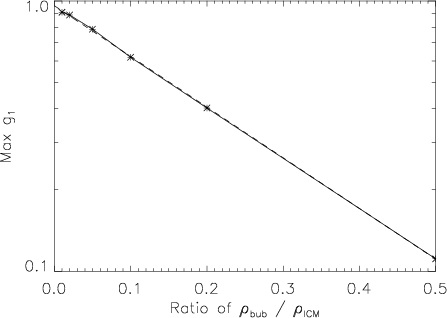}
  \caption{$g_1$ as a function of bubble density contrast
    ($\rho_{\text{bub}} / \rho_{\text{ICM}}$) for a Mach 2
    shock. Dashed line is the fit to the data as expressed in equation
    \ref{eq:g_vs_bubble_dens}.}
\label{fig:g_vs_bubble_density_contrast}
\end{figure}

Since the efficiency of the RMI must approach zero as the density
contrast approaches unity, we expect $g_1$ to decrease as $\rho_{\rm
  bub}/\rho_{\rm ICM}$ increases.  In order to quantify this decrease,
we ran a set of simulations spanning a range of $10^{-2} \leq
\rho_{\rm bub}/\rho_{\rm ICM} \leq 0.5$.  We plotted the resulting
dependence of $g_1$ on the density contrast in
Fig.~\ref{fig:g_vs_bubble_density_contrast}.  Given the upper limits
on the filling factor of thermal gas inside cavities by
\cite{sanders:07}, the effect of mixing in cluster will, at most,
reduce the effect of the RMI on cavities by $\sim$30\%.  The
dependence of $g$ on $\rho_{\rm buble}/\rho_{\rm ICM}$ is well fit by
the ad-hoc expression
\begin{align}
  \label{eq:g_vs_bubble_dens}
  g_1 &= g_{\infty} \times e^{-4.32 \times 
    \rho_{\text{bubble}}/\rho_{\text{ICM}}} & g_{\infty} &= 0.960
\end{align}

\subsection{Placing This Investigation into a Broader Fluid Mechanics
  Context}
\label{sec:calc_vortex_ring_vels}

We can derive a number of easily measurable quantities from our set of
simulations that allow comparison to other studies in the fluid
mechanics literature and are useful experimental diagnostics of the
RMI.

\subsubsection{Vortex velocities}
When a shock encounters a bubble it creates at least one, often two,
vortex rings.  The upstream vortex, which contains most of the RKE,
always forms and, following literature convention, we refer to this
vortex as the primary vortex ring (PVR).  A \downstream vortex
sometimes forms with a usually smaller radius, which we refer to as
the secondary vortex ring (SVR).  We can clearly see these vortex
rings in Figure~\ref{fig:dens_slice_0030_nz-2.jpg}; at that time for
the Mach 2 case, the SVR has just started to form, as noted by the
curl-up downstream from the PVR.

We note that a SVR does not always occur in a sub-set of our
simulations, in particular:
\begin{itemize}
\item{at low Mach numbers ($M \lesssim 1.07$), mainly
  because it takes so long for vortices to form}
\item{in the case of multiple bubbles}
\item{in simulations of cylinders with axes parallel to the shock
  normal}
\item{at high Mach numbers (e.g.~Mach 8).}
\end{itemize}

\begin{figure}[tpb]
  \centering
  \includegraphics[width=0.95\columnwidth]{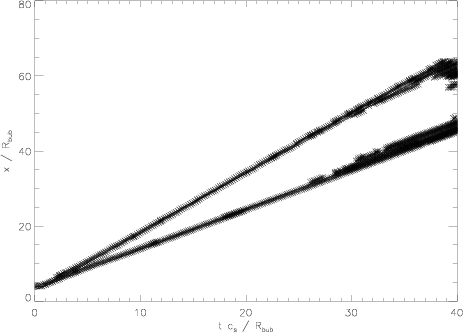}
  \caption{Location of maxima in fluid tracer distribution vs.~time
    for our fiducial Mach 2 run.  Only those maxima with a maximum
    value with 3\% of the global maximum for each time are plotted.
    The top line corresponds to the movement of the PVR, whereas the
    bottom line corresponds to the movement of the SVR.}
  \label{fig:in_maxima_plot_mach2-3.ps}
\end{figure}

We track the locations and velocities of both PVR and SVR by finding
the peak values of the tracer fluid (projected onto the vortex axis).
Figure~\ref{fig:in_maxima_plot_mach2-3.ps} shows the resulting
locations for our fiducial Mach 2 simulation, indicating that the the
SVR slowly drifts away from the PVR and that both move at near
constant velocity.  

\begin{figure}[tpb]
  \centering
  \includegraphics[width=0.95\columnwidth]{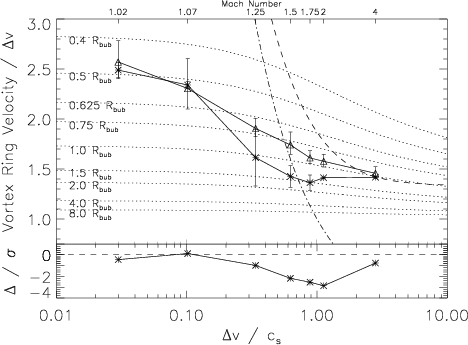}
  \caption{Measured velocity of the primary vortex ring (triangles) as a
    function of Mach number, compared to the prediction from
    eq.~(\ref{eqn:picone_boris_vortex_vel_our_notation}) and the
    measured vortex radii $R_{\rm PVR}$ (stars).  For comparison,
    the dashed line represents the speed of the shock front for a
    given $\Delta v$, whereas the the dashed-dotted line represents
    the \upstream sound speed, $c_s$.  The bottom portion of the plot
    represents the difference between the two lines divided by the
    uncertainty of both curves added in
    quadrature.\label{fig:pvr_sph_errors_gray}}
\end{figure}

In Fig.~\ref{fig:pvr_sph_errors_gray} we plot the best fit PVR
velocities from linear regressions to the positions of the tracer
maxima (the variance in $v_{\rm PVR}$ is reflected in the error bars).
It is clear from the figure that the PVR generally slows down relative
to the \downstream velocity as the Mach number increases.

\citet[\S VI]{2008PhFl...20c6101R} provide an analytic estimate of the
velocity of the PVR as a function of the vortex ring radius $R_{\rm
  PVR}$, based on the model from \citet{1988JFM...189...23P}.
\begin{align}
  \label{eqn:picone_boris_vortex_vel_our_notation}
  \frac{v_{\text{PVR}}}{\Delta v} &= 1 + \frac{R_{\text{PVR}}}{2 \pi
    D_V} \left(1 - \frac{\Delta v}{2 M c_s} \right) \ln
  \left( \frac{\rho_{\text{ICM}}}{\rho_{\text{bub}}} \right)
\end{align}
To compare our work to prior results \citep{2008PhFl...20c6101R}, we
converted their data using
eq.~\ref{eqn:picone_boris_vortex_vel_our_notation} for a fluid with
adiabatic index $\gamma = 5/3$ and a density contrast of
$\rho_{\text{ICM}}/\rho_{\text{bub}} = 100$ and measured radii $R_{\rm
  PVR}$ for the vortex rings for different Mach numbers.

This model provides a reasonable approximation for the PVR velocities
for low Mach numbers, as shown in Fig.~\ref{fig:pvr_sph_errors_gray}.
The top panel of the figure shows the measured PVR velocity
(triangles) and the predicted values from
eq.~(\ref{eqn:picone_boris_vortex_vel_our_notation}).

To calculate the errors in $v_{\rm PVR}$, we selected points that
belonged to the PVR in a figure like
Fig.~\ref{fig:in_maxima_plot_mach2-3.ps} and performed a linear
regression of those points.  From the fit, we obtain a slope and an
error for our fit.  To measure the ring radii, we performed a weighted
average on the tracer fluid.  The variance in these measurements is
reflected in the error bars in the stars in
Fig.~\ref{fig:pvr_sph_errors_gray}.

In the lower panel of Fig.~\ref{fig:pvr_sph_errors_gray}, we plot the
difference between the model and the measurements for $v_{\rm PVR}$
($\Delta$) divided by the estimated uncertainty ($\sigma$).

\subsubsection{Vortex turnover locations and times}

The {\em time} it takes for the vortex to form after the shock first
encounters the bubble provides an important diagnostic, both
astrophysically and in laboratory experiments.

The {\em distance} traveled by the shock at the time the vortex forms
--- the minimum stand-off distance between vortex and shock ---
provides another diagnostic that can be applied directly to
astrophysical observations.  The stand-off distance is also a direct
measure of the vortex velocity relative to the shock velocity in the
downstream frame.

The criterion for vortex formation in the simulations is most easily
defined as the condition that a spine of high-density, shocked
material passes all the way through the bubble.  Technically, this
criterion is satisfied when no more tracer fluid can be found along
the axis.

\begin{figure}[tpb]
  \centering
  \includegraphics[width=0.95\columnwidth]{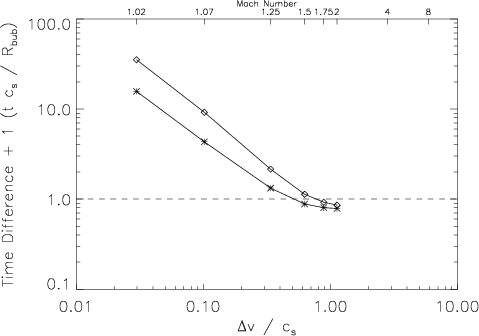}
  \caption{Time elapsed between when shock front passes location where
    vortex will form and when the vortex forms. The dashed horizontal
    line represents when the shock front reaches the location of
    vortex formation at the time the vortex forms. Stars represent
    spheres and diamonds represent
    cylinders.  \label{fig:turnover_plot_time_110209.ps}}
\end{figure}

The turnover time is shown in
Figure~\ref{fig:turnover_plot_time_110209.ps}, measured in units of
shock crossing times of the bubble.

\begin{figure}[tpb]
  \centering
  \includegraphics[width=0.95\columnwidth]{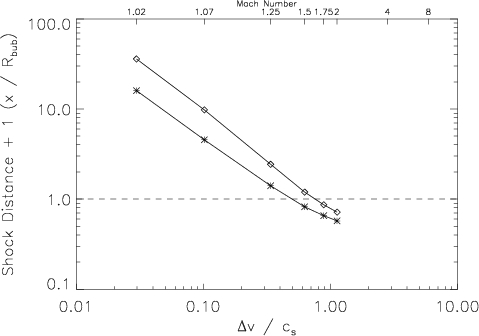}
  \caption{Plot of the distance between shock front and vortex at the
    time of vortex formation.  The dashed horizontal line represents
    when the shock front reaches the location of vortex formation at
    the time the vortex forms.  Stars represent spheres and diamonds
    represent cylinders.  Since the vortex, has to travel more slowly
    than the shock, this is the minimum distance that can be observed
    between a shock and a
    vortex.\label{fig:turnover_plot_distance_110209.ps}}
\end{figure}

A plot of the stand-off distance as a function of Mach number is shown
in Fig.~\ref{fig:turnover_plot_distance_110209.ps}, in units of
initial bubble radii.  Given an observational estimate of the
stand-off distance, it is possible to derive a lower limit on the Mach
number from this figure, since the minimum stand-off distance
decreases with increasing Mach number.

\section{Dissipation of wave energy in the intracluster medium}
\label{sec:dissipation}

As originally suggested in HC05, the kinetic energy contained in the
vortex field generated in the wake of a shock or steepened sound wave
passing over filaments of hot/relativistic gas in galaxy cluster cores
might be dissipated on time scales long compared to the shock passage
time but short compared to the cluster cooling time.  The presence of
multiple generations of cavities of relativistic plasma in the cluster
could thus enhance the dissipation of acoustic energy released by the
generation of subsequent cavities through the activity of the AGN.

A vortex created by this process exhibits a differentially rotating
velocity profile, which can be seen from the radial decline in the
vortex energy density in Fig.~\ref{fig:dens_slice_0030_nz-2.jpg}.  In
the presence of microscopic or turbulent viscosity, the induced strain
in the velocity field will ultimately lead to dissipation of the
kinetic energy in the vortex.

Based on the results of our 3D parameter study of the efficiency of
the RMI, we can estimate the dissipation rate for the vortex energy
and determine under which conditions we might expect it to contribute
significantly to the thermodynamics of the cluster gas.

\subsection{Viscous Dissipation of Vortex Rings}

As we discussed in \S\ref{sec:high_mach_numbers}, {\em strong} shocks
generate vortices that are inherently dynamically unstable on a shock
passage timescale.  The vortex field generated by the shock thus
dissipates in a turbulent cascade quickly after the shock passed, with
high efficiency.  Strong shocks are also {\em inherently} dissipative,
and any region of the cluster subject to such a shock will be heated
efficiently regardless of the presence or absence of the RMI.

However, given the well known evolution of expanding AGN driven
cavities in clusters \citep[e.g.][]{heinz:98}, only a small volume and
mass fraction of the cluster will be subject to such strong shocks,
while most of the cluster only experiences relatively weak shocks,
consistent with the observations of shock Mach numbers in the range of
1-2 in clusters where shocks have been discovered.

As demonstrated above, the vortices generated for such weak to
moderate shocks are dynamically stable for times much longer than the
shock crossing time, and the shock or sound wave itself will not
contribute sufficiently to the heating of the gas to offset cooling
unless the viscosity is close to the Spitzer value
\citep{2005MNRAS.357..242R}.

However, the vortex ring itself is differentially rotating.  As
originally suggested in HC05, viscous
dissipation due to the shear in this flow will transfer some of the
rotational kinetic energy in the vortex to heat in the cluster gas on
a viscous dissipation time scale $\tau_{\rm diss}$.

It is easy to derive the natural scaling of $\tau_{\rm diss}$ with
vortex parameters: Following equations~\ref{eqn:fidual_energy} and
\ref{eqn:define_g}, the vortex energy is given by $E_{\rm rot}=V_{\rm
  bubble}\rho_{2}\frac{(\Delta v)^2}{2}g_{1}$.

The characteristic velocity of the vortex is simply $\Delta v$; the
velocity decreases from $\sim\Delta v$ outward from the vortex
surface.  It is clear that $\Delta v$ must be the velocity scale
imposed by the initial conditions and our simulations confirm this
(see \citet{1988JFM...189...23P,batchelor} for a more rigorous
motivation).

The characteristic volume inside which most of the vortex is contained
must be of the order of $V_{\rm bubble}$.  Finally, the vortex must
have a characteristic scale length of the order of the initial bubble
radius, $r_{\text{bub}}$.

Consequently, the shear inside this volume is of the order of
\begin{equation}
  \frac{\partial v_{\rm rot}}{\partial r} \sim \frac{\Delta v}{r_{\rm bubble}}
\end{equation}

Following \citep{2005MNRAS.357..242R}, we write the viscosity in terms
of the Spitzer-Braginsky value
\citep{1958JETP....6..358B,1962pfig.book.....S}
\begin{align}
  \mu &= 2.21 \times 10^{-15} \frac{T^{5/2}}{Z^4 \ln \Lambda}
  \frac{\text{g}}{\text{cm s}} \\ \nonumber &= 1.88 \times 10^3
  T_5^{5/2} \text{g cm}^{-1} \text{ s}^{-1}
\end{align}
where we have introduced a fiducial cluster temperature of $kT = 5 T_5
\text{ keV}$ and used $\ln \Lambda \approx 30$.

We also define a fractional viscosity parameter (i.e., the viscosity
measured in units of the Spitzer value) as
\begin{equation}
  f_{\rm Sp} \equiv \frac{\mu}{\mu_{\rm Spizter}}
\end{equation}

The dissipation rate for a vortex with these characteristic parameters
will then be of the order of\footnote{The first line in
  eq.~\ref{eq:diss1} states that the dissipation rate is set by the
  contraction (denoted by a colon) of the viscous stress tensor $\Pi$
  with the strain tensor $\nabla \vec{v}$}
\begin{eqnarray}
  \left.\frac{dE}{dt}\right|_{\rm diss} & = &  \int dV \Pi : \nabla
  \vec{v} \label{eq:diss1} \\ 
  & = & \int dV \mu \left(\frac{\partial v_{i}}{\partial x_{j}} +
    \frac{\partial v_{j}}{\partial x_{i}} - \frac{2}{3}\nabla\cdot
    \vec{v}\right) \\\nonumber & \quad & \times \left(\frac{\partial
      v_{i}}{\partial v_{j}} +
    \frac{\partial v_{j}}{\partial x_{i}}\right) \label{eq:dissipation}\\
  & \equiv & 2 V_{\rm bubble} \mu \left(\frac{\Delta
      v}{R}\right)^2\xi_{\rm diss}
\end{eqnarray}
where we introduced the dissipation efficiency coefficient $\xi_{\rm
  diss}$, to be evaluated from the actual shear and vortex volume
measured in the simulation.

The dissipation time, using eqs.~\ref{eqn:fidual_energy} and
\ref{eqn:define_g}, will then be of the order of
\begin{eqnarray}
  \tau_{\rm diss} &\equiv& \frac{E_{\rm rot}}{dE/dt} \\
  & \sim & \frac{V_{\rm
      bubble}\rho_{2}\frac{(\Delta v)^2}{2}g_{1}}{\mu
    \frac{(\Delta v)^2}{R^2}V_{\rm bubble}}\xi_{\rm diss}^{-1} =
  \frac{R_{\text{bub}}^2 g_{1}
    \rho_{2}}{2\mu \xi_{\rm diss}}
  \label{eq:t_diss_expressed}
  \\
  & \sim & 1.3\times 10^{6}\,{\rm yrs}\,\frac{R_{\rm
      kpc}^2\rho_{0.01}g_{1}}{f_{\rm Sp} T_{5}^{5/2}\xi_{\rm diss}}
  \label{eq:t_diss_expanded}
\end{eqnarray}
which we use as our fiducial reference scale to plot the numerically
determined dissipation rates against.

While our simulations were inviscid (with the exception of numerical
viscosity and artificial viscosity employed in the shock-capturing
scheme), we can calculate the viscous dissipation rate and thus
$\xi_{\rm diss}$ a posteriori, using a finite-difference
representation of eq.~\ref{eq:dissipation}.

Our simulations approximate the relativistic, non-thermal plasma
inside the vortex as hot, thermal gas.  Given the steep temperature
dependence in $\mu$, and given the high temperatures inside the
vortex, care must be taken in excluding any contribution to the
dissipation rate from inside the vortex itself, which would be
unphysical.

To this end, we impose a temperature cutoff on the gas, motivated by
the fact that the post-shock gas around the vortex occupies a
relatively narrow range in temperature, clearly separate from the much
hotter vortex.  We chose a conservative cut of $T_{\rm
  cut}=1.25T_{2}$, which effectively excludes most of the hot vortex.

We performed this analysis on both the entire velocity field,
$\mathbf{v}$, and just the rotational component of the velocity field,
$\mathbf{v}_R$.  If the analysis is limited to exclude the shock
(which contributes to the viscous dissipation rate of the full
velocity field but is naturally absent in the rotational component of
the flow), we find that the late-time difference between the
dissipation rates for the rotational and the full velocity field is
less than 5\%.  This is consistent with the absence of any
significant viscous dissipation in the acoustic part of the velocity
field.

\begin{figure}[tpb]
  \centering
  \includegraphics[width=0.95\columnwidth]{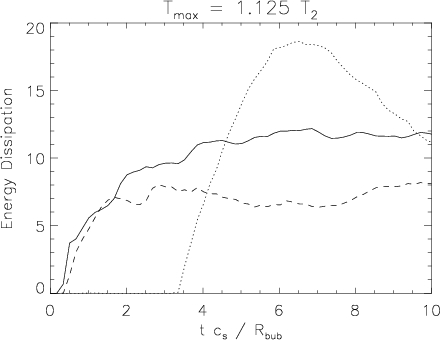}
  \caption{Viscous Energy Dissipation Rate, $\xi_{\text{diss}}$, in
    units of the fiducial dissipation rate from
    eq.~(\ref{eq:dissipation}), plotted against simulation time in
    units of bubble sound crossing times.  The solid line represents a
    Mach 2 shock interacting with a spherical bubble; the short dashed
    line corresponds to a Mach 1.5 shock, and the dotted line to a
    Mach 1.07 shock.}
  \label{fig:dissipation}
\end{figure}

\begin{figure}[tpb]
  \centering
  \includegraphics[width=0.95\columnwidth]{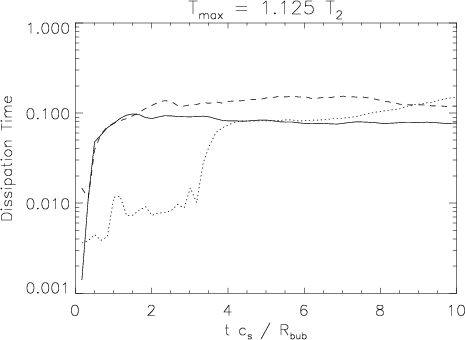}
  \caption{Viscous Energy Dissipation Time in units of the fiducial
    dissipation time from eq.~(\ref{eq:t_diss_expanded}), equal to the
    ratio of the dimensionless factors $g_1/\xi_{\text{diss}}$ (same
    line styles as Fig.~\ref{fig:dissipation}}
  \label{fig:tdiss}
\end{figure}

The viscous dissipation rate in units of $\left.dE/dt\right|_{\rm
  diss}$ from eq.~(\ref{eq:dissipation} for different Mach numbers as
a function of time is shown in Fig.~\ref{fig:dissipation}.  Figure
\ref{fig:tdiss} plots the inferred dissipation time in units of
$\tau_{\rm diss}$ from eq.~(\ref{eq:t_diss_expanded}).

As can be seen from Fig.~\ref{fig:dissipation}, the dissipation rate
inferred from strain and vortex volume measured in the simulations is
about an order of magnitude larger than the natural scaling derived in
eq.~\ref{eq:dissipation}, i.e., $\xi_{\rm diss} \sim 10$, and we will
use $\xi_{\rm diss}\equiv 10\xi_{10}$ as our fiducial value through
the rest of the discussion.  We attribute this to the fact that the
velocity gradients inside the vortex are, in fact, significantly
steeper and more concentrated than the simple estimate of $\Delta
v/r_{\rm bubble}$ would suggest (which we confirmed by inspection of
individual frames of the simulation).

This carries over to the estimated dissipation times of vortices in
clusters, which is also about an order of magnitude shorter than the
fiducial rate.

\subsection{Application to Galaxy Clusters}

The dissipation time should be compared to the residence time of the
vortex (i.e., the travel time through the cooling region) and the
cluster cooling time in order to assess the viability of this process
to contribute to the thermalization of AGN energy in clusters.
Technically, these depend on the cluster properties, but given that
cooling times in the centers of cool core clusters are of the order of
a few hundred million years.

Taking the viscous dissipation efficiency plotted in
Fig.~\ref{fig:tdiss}, and denoting the cooling function of the gas as
$\Lambda(T) \equiv 10^{-23}\,{\rm
  ergs\,cm^{3}\,s^{-1}}\Lambda_{-23}(T)$, the cooling time is longer
than the dissipation time if
\begin{equation}
  \frac{\Lambda_{-23}(T)\,R_{\rm kpc}^2\,\rho_{0.01}^2g_{1}}{f_{\rm
      Sp}T_{5}^{7/2}} < 3\times 10^{4}
  \label{eq:rdiss}
\end{equation}

If the dissipation time is long compared to the cooling time, an AGN
driven feedback loop will not be able to counteract the onset of
cooling rapidly enough to maintain thermal balance of the cluster.
Whether the condition $\tau_{\rm diss} < \tau_{\rm cool}$ evaluation
in equation~\ref{eq:rdiss} is satisfied (i.e., whether the onset of
AGN activity occurs at high enough temperatures for dissipation in the
IGM to be efficient) will depend on the details of gas supply to the
black hole as a function of central cluster
temperature\footnote{Molecular viscous dissipation shares this strong
  dependence on temperature with conduction as a heating agent in
  clusters.}, given the strong temperature sensitivity of
eq.~\ref{eq:rdiss}.  

Note, however, that the condition $\tau_{\rm diss} < \tau_{\rm cool}$
is not a strict requirement for feedback to work, as long as a
sufficient fraction of the vortex energy is dissipated in the cooling
region to offset cooling in an average sense, as eventually, a
sufficient amount of energy will be liberated by the AGN to counteract
cooling.

A more important requirement for effective heating is that the vortex
remain in the cluster core long enough to dissipate a significant
fraction of its energy.  The vortex residence time in the cluster is
more difficult to estimate than the dissipation time: Our simulations
are idealized in that they model the shock as a semi-infinite piston.
And shocks and non-linear sound waves in clusters are impulsive, and
thus the long term dynamics of the vortex might be different from our
idealized simulations.

With this caveat in mind, we conservatively use our estimates of the
primary vortex velocities from Fig.~\ref{fig:pvr_sph_errors_gray} to
derive a rough estimate of the residence time.  The figure shows that
the velocity of the primary vortex ring (which contains the bulk of
the vortex energy) travels at velocities between 1.4 and 2.5 times the
velocity differential $\Delta v$, with vortices produced by weaker
shocks traveling relatively faster compared to $\Delta v$.

While for strong shocks this implies that the vortex travels close to
the shock speed (consistent with the fact that the standoff distance
between vortex and shock can be very small, as seen in
Fig.~\ref{fig:turnover_plot_distance_110209.ps}), for weaker shocks,
the vortex travels significantly more slowly than the shock, which
travels essentially at the sound speed, while $\Delta v$ is much
smaller than $c_{\rm s}$.

From Fig.~\ref{fig:pvr_sph_errors_gray}, we can see that the vortex
velocity is smaller than the sound speed of the cluster for shocks
with Mach number below about 1.6, with vortices created by weak shocks
traveling at very sub-sonic speeds.  For most of its propagation
through a cluster, an AGN driven shock will be below this critical
Mach number.

Taking the conservative upper limit on the PVR velocity to be about
$v_{\rm PVR} \lesssim 2.5\Delta v$, the actual velocity through the
ICM will be
\begin{align}
  \frac{v_{\rm PVR}}{c_{\rm s}} & \lesssim 2.5 \frac{\Delta v}{c_{\rm
      s}} = 2.5 \times \frac{3}{4}\left(M - \frac{1}{M}\right)
\end{align}
where $c_{\text{s}}$ is the \upstream sound speed and we have used
$\gamma = 5/3$ in calculating $\Delta v$ as a function of $M$.

For a cluster with cooling radius $r_{\rm cool} \equiv 50\,{\rm
  kpc} \, r_{50}$, the travel time through the cooling region is then
\begin{align}
  \tau_{\rm travel} &\sim \frac{r_{\text{cool}}}{v_{\text{PVR}}} \sim
  \frac{r_{\rm cool}}{2.5 c_{\rm s}}\frac{4}{3}\frac{M}{M^2 - 1}
  \\\nonumber &\sim 2\times 10^{7}\,{\rm
    yrs}\frac{r_{50}}{T_{5}^{1/2}}\frac{M}{M^2 - 1}
\end{align}
and the dissipation time is smaller than the travel time if
\begin{align}
  \frac{R_{\rm kpc}^2\rho_{0.01}g_{1}}{f_{\rm Sp}T_{5}^2 r_{50}
    \xi_{10}}\left(M - 1/M\right) \lesssim 150
\end{align}

We can compare the viscous dissipation time to the eddy turnover time
(i.e., the turbulent dissipation time),
\begin{equation}
  \tau_{\rm turb} \sim \frac{R_{\rm bub}}{\Delta v} =
  7\times 10^{6}\,{\rm yrs}\frac{R_{\rm kpc}}{T_{5}^{1/2}\left(M_1 -
      1/M_1\right)}
\end{equation}
which is shorter than the viscous dissipation time if
\begin{equation}
  \frac{f_{\rm Sp} T_{5}^2 \xi_{10}}{R_{\rm kpc} \rho_{0.01}
    g_{1} \left(M_1 - 1/M_1\right)}
  \lesssim 0.02
\end{equation}
which, not surprisingly, is the case for large bubbles and strong
shocks.

Whether the dissipation of vortex energy contributes significantly to
cluster heating will, of course, ultimately depend on the energy
released by the AGN.  Even if the dissipation time is short compared
to the travel time and the cooling time, a sufficient amount of energy
has to be injected into waves and then extracted into the vortex
field, which depends on the filling factor of cavities in clusters and
the on AGN energy output relative to the cooling rate.  

As already shown in HC05, the attenuation
length of a wave with width $\lambda \gg R_{\rm bubb}$ interacting
with a field of underdense bubbles of filling factor $f$ is roughly
\begin{equation}
  L \sim \frac{\lambda}{2g_{1}f}
\end{equation}
where we have included the factor 2 increase in efficiency introduced
by extending the analysis to 3D.  For the wave to lose most of its
energy within the cooling radius, the filling factor would have to be
larger than
\begin{equation}
  f \gtrsim \frac{10\%}{g_{1}} \left( \frac{50\,{\rm
        kpc}}{r_{\rm cool}}\frac{\lambda}{10\,{\rm kpc}} \right)
\end{equation}

Given that all cool core clusters show clear evidence of bubbles, and
in cases where statistics allow, multiple generations thereof, such
large filling factors are not unreasonable and consistent with
estimates of the amount of non-thermal pressure present in the ICM of
nearby clusters.

While the uncertainty in the relevant parameters (namely, the
distribution of $R_{\rm bubb}$ of radio plasma, and the filling factor
$f$, the Spitzer fraction $f_{\rm SP}$) is too large to conclude that
visco-rotational heating is an important contributor to the thermal
evolution, the study shows that the process should be studied further:
It is clear that it {\em must} happen at some level when AGN-driven
waves pass over the existing pockets of radio plasma, and under the
right conditions, it can be very important in the extraction of energy
from sound and shock waves in clusters.

\section{Conclusions}
\label{sec:conclusions}
We presented a detailed numerical investigation of the efficiency of
the Richtmyer-Meshkov instability in the context of shocks passing
over radio plasma filaments and cavities in galaxy cluster
atmospheres.  We investigated the possibility that extraction and
dissipation of energy from weak shocks and non-linear sound waves
often found in the centers of cool core clusters could contribute
significantly to the heating of cool core clusters.

We introduced a 3D solenoidal/Helmholtz decomposition as an analytic
tool to study the efficiency of vortex generation and to quantify the
energy deposited in the vortex field upon passage of the shock over a
bubble.

We generally confirmed the previous calculations of
HC05 in the 2D limit and extended the
analysis to full 3D simulations.  We found that, for roughly spherical
bubbles, the efficiency of vortex generation, as measured by the
kinetic energy in the vortex field, is increased by a factor of 2 over
the 2D case.

In the case of high Mach numbers ($M \gtrsim 4$), we found that the
simulated vortices are not stable and degenerate into turbulence.  The
vortex energy is quickly dissipated and the vortex shredded (this can
be clearly seen from
Figure~\ref{fig:bw_3d_mach8_bis_hdf5_plt_cnt_in0060.jpg}).  Generally,
strong shocks are not observed in the centers of clusters ICM, so we
should expect RMI generated vortices in cluster atmospheres to be
stable, though this might not be the case in the very centers of
clusters in the presence of powerful, young radio galaxies.  Lastly,
strong shocks will raise the temperature of the gas on their own can,
eliminating the requirement for additional dissipation mechanisms to
facilitate AGN energy deposition in response to cluster cooling.

We found that non-linear interactions of multiple vortices can
dynamically disrupt the vortices, leading to enhanced dissipation and
a rapid decline in the kinetic energy in the vortex field, similar to
what is seen in the case of large Mach numbers.  While further
investigation of this effect is necessary, we speculate that this is
similarly due to the development of turbulence, as was found in the 2D
case of random two-phase gas distributions in
HC05.  As a result of this effect, the
efficiency of vortex generation and dissipation will depend on the
average distance between vortices, i.e., the volume filling factor of
low density plasma, with values significantly in excess of a few \%
indicating strongly non-linear interaction between vortices.

Since the vortex is a differentially rotating flow, it must be subject
to viscous dissipation.  We found that viscous dissipation of the
vortex is about an order of magnitude more efficient than would be
expected from a simple dimensional scaling argument.  We showed that
the viscous dissipation time is shorter for smaller vortices (i.e.,
bubbles) and that, in general, it can be smaller than the cluster
crossing time and the ICM cooling time for bubbles smaller than about
10 kpc if the viscosity is at the few percent level of the Spitzer
value.  Observations of the center of nearby clusters
\citep{2005MNRAS.360L..20F,2005ApJ...635..894F,2002ApJ...579..560Y}
indicate that the multi-phase gas in cluster centers might have the
right properties for viscous dissipation to contribute to cluster
heating.

For this process to be thermodynamically relevant to the ICM within
the cooling radius, the filling factor of non-thermal plasma must be
significant (again of the order of a few percent).  An investigation
of how the non-linear interaction of closely spaced vortices will
affect the heating of the ICM via the RMI is beyond the scope of this
paper.

In order to compare our results to the existing experimental and
numerical body of work on the RMI, we examined the velocities of the
vortex rings in our simulations\footnote{The vortex travel velocities
  are also important in determining timing and location of energy
  dissipation in the ICM.}.  We found that the velocities of the
primary vortex ring (PVR) roughly match those predicted by
\citet{1988JFM...189...23P} and expressed in
equation~\ref{eqn:picone_boris_vortex_vel_our_notation}, with some
significant deviations at intermediate Mach numbers\footnote{this is
  not entirely surprising, given that the original formula is based on
  analytic considerations and 2D simulations.}.

Finally, we introduced a new diagnostic for constraining the Mach
number of shock waves present in the ICM: The stand-off distance
between shock and vortex (see
Fig.~\ref{fig:turnover_plot_distance_110209.ps}). Since vortex
formation is slower, relative to shock passage, for weaker shocks, an
observed stand-off distance can provide a lower limit on the Mach
number.

\subsection{Acknowledgements}
We would like to thank Mateusz Ruszkowski, Marcus Br{\"u}ggen, Ellen
Zweibel, Rich Townsend, Eric Wilcots, and Riccardo Bonazza for helpful
comments and discussions.  The software used in this work was in part
developed by the DOE-supported ASC / Alliance Center for Astrophysical
Thermonuclear Flashes at the University of Chicago.  Thank you to the
CHTC for computational resources.  SH and SHF acknowledge support from
NASA through Chandra theory grant TM9-0007X and from NSF through grant
AST0707682.

\end{document}